\documentclass[journal]{IEEEtran}
\IEEEoverridecommandlockouts
\usepackage{lineno}
\usepackage{cite}
\usepackage{hyperref}
\usepackage{amsmath,amssymb,amsfonts}
\usepackage{amsmath}
\usepackage{amsthm}

\usepackage{graphicx}
\usepackage{textcomp}
\usepackage{xcolor}
\usepackage{graphicx}
\usepackage{float}
\usepackage{subfigure}
\usepackage{amsmath}
\usepackage{amsfonts,amssymb}
\usepackage{mathrsfs}
\usepackage{mathtools}
\usepackage{algorithm}
\usepackage{algorithmicx}
\usepackage{algpseudocode}
\usepackage{bm}
\usepackage{multirow}
\usepackage{array}
\usepackage{amssymb}
\usepackage{amsmath}
\usepackage{cite}
\usepackage{url}
\usepackage{xcolor}
\usepackage{cite,graphicx,amsmath,amssymb}
\usepackage{subfigure}
\usepackage{fancyhdr}
\usepackage{mdwmath}
\usepackage{mdwtab}
\usepackage{caption}
\usepackage{amsthm}
\usepackage{setspace}
\usepackage{bm}
\usepackage{algorithm}
\usepackage{algpseudocode}
\usepackage{mathtools}
\usepackage{dsfont}
\usepackage{bbm}
\newtheorem{remark}{Remark}
\newtheorem{theorem}{Theorem}

\newtheorem{lemma}{Lemma}

\newtheorem{corollary}{Corollary}

\makeatletter
\newcommand{\biggg}{\bBigg@{3}}
\newcommand{\Biggg}{\bBigg@{3.5}}
\makeatother
\begin{document}
\title{NOMA-ISAC: Performance Analysis and Rate Region Characterization}
\author{Chongjun~Ouyang, Yuanwei~Liu, and Hongwen~Yang
\thanks{C. Ouyang and H. Yang are with the School of Information and Communication Engineering, Beijing University of Posts and Telecommunications, Beijing, 100876, China (e-mail: \{DragonAim,yanghong\}@bupt.edu.cn).}
\thanks{Y. Liu is with the School of Electronic Engineering and Computer Science, Queen Mary University of London, London, E1 4NS, U.K. (e-mail: yuanwei.liu@qmul.ac.uk).}
}
\maketitle

\begin{abstract}
This paper analyzes the performance of a multiuser integrated sensing and communications (ISAC) system, where nonorthogonal multiple access (NOMA) is exploited to mitigate inter-user interference. Closed-form expressions are derived to evaluate the outage probability, ergodic communication rate, and sensing rate. Furthermore, asymptotic analyses are carried out to unveil diversity orders and high signal-to-noise ratio (SNR) slopes of the considered NOMA-ISAC system. As the further advance, the achievable sensing-communication rate region of ISAC is characterized. It is proved that ISAC system is capable of achieving a larger rate region than the conventional frequency-division sensing and communications (FDSAC) system.
\end{abstract}


\section{Introduction}
Next-generation wireless networks are envisioned to provide the capability of simultaneous communications and sensing \cite{Liu2022}. This gave birth to the concept of integrated sensing and communications (ISAC), where wireless communications and radar sensing are integrated together to share the same spectrum and hardware platform \cite{Liu2022}. Compared with the conventional frequency-division sensing and communications (FDSAC) scheme, in which sensing and communications require isolated frequency bands as well as hardware infrastructures, ISAC is expected to be more spectrum- and hardware-efficient and has sparked a significant burst of recent research attention \cite{Liu2022,Liu2021,Wang2021,Mu2021,Liu2022_2,Chiriyath2016,Rong2018,Yuan2021}.

From an information-theoretic perspective, the performance of ISAC can be evaluated by two fundamental metrics, i.e., sensing rate (SR) and communication rate (CR) \cite{Liu2021}. Specifically, SR measures how much environmental information can be estimated, wheras CR measures how much data information can be transmitted \cite{Liu2021}. These two metrics were discussed in several studies for the sake of demonstrating the information-theoretic limits of ISAC systems; see \cite{Liu2021,Chiriyath2016,Rong2018,Yuan2021} and the references therein. However, we notice that these studies would have made more practical sense if the authors had took account of the influence of both inter-user interference (IUI) and channel fading. Furthermore, despite its significance, a rigorous comparison between the sensing-communication rate regions achieved by ISAC and FDSAC is still missing.

In this letter, we investigate the performance of a multiuser ISAC system where a nonorthogonal multiple access (NOMA) protocol is involved in the communication procedure for IUI mitigation \cite{Liu2017,Liu2023}. For the sake of convenience, this system is also termed ``NOMA-ISAC''. It is worth mentioning that the NOMA-ISAC framework was firstly established in \cite{Wang2021}. Yet, this work focused more on the waveform design and shed less light on the fundamental properties of SR as well as CR. The main contributions of our work are summarized as follows. Firstly, by taking the influence of IUI as well as channel fading into consideration, we derive closed-form expressions for outage probability (OP), ergodic CR (ECR), and SR, respectively. Secondly, by letting the signal-to-noise ratio (SNR) go to infinity, we perform asymptotic analyses to the aforementioned metrics in order to unveil the diversity order as well as high-SNR slope of the NOMA-ISAC system. Finally, based on the developed analytical results, we characterize the sensing-communication rate region achieved by NOMA-ISAC and prove that ISAC is capable of achieving a larger rate region that FDSAC.

\begin{figure}[!t]
\centering
\setlength{\abovecaptionskip}{0pt}
\includegraphics[height=0.25\textwidth]{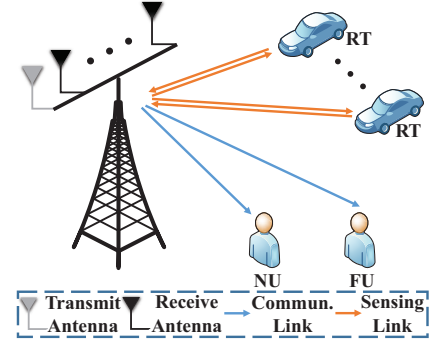}
\caption{Illustration of an NOMA-ISAC system}
\label{System_Model}
\vspace{-20pt}
\end{figure}

\section{System Model}
There is a radar-communications (RadCom) base station (BS) equipped with a single transmit antenna and $M$ receive antennas, which is serving a pair of single-antenna communication users (CUs)\footnote{The developed analytical results can be extended to systems with more than two CUs. We skip further details for sake of brevity.}, i.e., one near user (NU) and one far user (FU), while sensing the radar targets (RTs) in the surrounding environment, as depicted in {\figurename} {\ref{System_Model}}. Let $\mathbf{x}=\left[x_1,\cdots,x_L\right]^{\mathsf{T}}\in{\mathbbmss{C}}^{L\times 1}$ be a dual-functional signal vector, with $L$ being the length of the communication frame/radar pulse. From a sensing perspective, $x_l$ for $l\in{\mathcal{L}}=\left\{1,\cdots,L\right\}$ represents the radar snapshot transmitted at the $l$th time slot. For communications, $x_l$ is the $l$th data symbol.
\subsection{Communication Model}
By transmitting $\mathbf{x}$ to the two CUs, the received signal matrix at the CU pair is given by
\begin{align}
{\mathbf{X}}_{\rm{c}}={\mathbf{h}}{{\mathbf{x}}^{\mathsf{T}}}+{\mathbf{N}}_{\rm{c}}\in{\mathbbmss{C}}^{2\times L},
\end{align}
where ${\mathbf{N}}_{\rm{c}}\in{\mathbbmss{C}}^{2\times L}$ is the additive white Gaussian noise (AWGN) with variance of each entry being $\sigma_{\rm{c}}^2$, and ${\mathbf{h}}\in{\mathbbmss{C}}^{2\times1}$ is the communication channel matrix, which is assumed to be known by the BS. In this work, NOMA is employed at the BS for serving multiple CUs and the communication channel matrix reads ${\mathbf{h}}=\left[h_{\rm{N}},h_{\rm{F}}\right]^{\mathsf{T}}$, where $h_{\rm{N}}$ and $h_{\rm{F}}$ denote the channel coefficients of NU and FU, respectively. We assume that all the communication links shown in {\figurename} {\ref{System_Model}} suffer Rayleigh fading and the CUs are ordered based on their channel quality, i.e., $\left|h_{\rm{N}}\right|>\left|h_{\rm{F}}\right|$ \cite{Liu2017,Liu2023}. Consequently, we have $\left|h_{\rm{N}}\right|=\max\{|h_1|,|h_2|\}$ and $\left|h_{\rm{F}}\right|=\min\{|h_1|,|h_2|\}$, where $h_1\sim{\mathcal{CN}}\left(0,\varrho_1\right)$ and $h_2\sim{\mathcal{CN}}\left(0,\varrho_2\right)$ are two mutually independent Gaussian random variables with $\varrho_1>0$ and $\varrho_2>0$ indicating the fading severity. Under the framework of NOMA-ISAC, the signal vector $\mathbf{x}$ is given by
\begin{align}\label{dual_function_signal_matrix}
{\mathbf{x}}^{\mathsf{T}}=\sqrt{p}\left[\sqrt{\alpha_{\rm{N}}},\sqrt{\alpha_{\rm{F}}}\right]{\mathbf{S}},
\end{align}
where $p$ is the power budget at each time slot, $\alpha_u$ for $u\in\left\{\rm{N},\rm{F}\right\}$ denote power allocation factors with $\alpha_{\rm{N}}+\alpha_{\rm{F}}=1$ and $\alpha_{\rm{N}}<\alpha_{\rm{F}}$, and $\mathbf{S}\in{\mathbbmss{C}}^{2\times L}$ contains two unit-power data streams intended for the two CUs. The data streams are assumed to be independent with each other so that \cite{Liu2022_2,Wang2021,Mu2021}
\begin{align}\label{normalized_signal_matrix}
{\mathbf{S}}{\mathbf{S}}^{\mathsf{H}}\approx L\mathbf{I}_2.
\end{align}

\subsection{Sensing Model}
By transmitting $\mathbf{x}$ to sense the RTs, the reflected echo signal matrix at the receiver of the RadCom BS is given by
\begin{align}\label{reflected_echo_signal_matrix}
{\mathbf{Y}}_{\rm{s}}={\mathbf{g}}{\mathbf{x}}^{\mathsf{T}}+{\mathbf{N}}_{\rm{s}}\in{\mathbbmss{C}}^{M\times L},
\end{align}
where ${\mathbf{N}}_{\rm{s}}\in{\mathbbmss{C}}^{M\times L}$ is the AWGN matrix with variance of each entry being $\sigma_{\rm{s}}^2$, and $\mathbf{g}=\left[g_1,\cdots,g_M\right]^{\mathsf{T}}\in{\mathbbmss{C}}^{M\times1}$ represents the target response vector with $g_m$ for $m\in{\mathcal{M}}=\left\{1,\cdots,M\right\}$ representing the target response from the transmit antenna to the $m$th receive antenna. The target response vector can be written as \cite{Tang2019,Liu2022_2}
\begin{align}\label{target_response_vector}
{\mathbf{g}}=\sum\nolimits_{k}\beta_{k}{\mathbf{a}}\left(\theta_k\right),
\end{align}
where $\beta_{k}\sim{\mathcal{CN}}\left(0,\sigma_k^2\right)$ is the complex amplitude of the $k$th RT with $\sigma_k^2$ representing the average strength, ${\mathbf{a}}\left(\theta_k\right)\in{\mathbbmss{C}}^{M\times 1}$ is the associated receive array steering vector, and $\theta_k$ is its angle of arrival (AoA). It is worth noting that the BS has no prior knowledge about the number and the corresponding AoAs of RTs. Therefore, in essence, sensing the RTs is equivalent to extracting environmental information contained in $\mathbf{g}$, e.g., the AoA and reflection coefficient of each RT, from the reflected echo signal presented in \eqref{reflected_echo_signal_matrix} \cite{Liu2022,Liu2022_2}.

\subsection{Baseline Scheme}
In this letter, we consider NOMA-FDSAC as a baseline scenario, where the total bandwidth is partitioned into two sub-bands according to some $\kappa\in\left[0,1\right]$, one for radar only and the other for communications. Particularly, we assume $\kappa$ fraction of the total bandwidth is used for communications with the NOMA CU pair. In addition to this, the total power in FDSAC should also be partitioned into two parts according to some $\mu\in\left[0,1\right]$, one for radar only and the other for communications. Specifically, the powers for communications and sensing are given by $\mu p$ and $\left(1-\mu\right)p$, respectively. Moreover, it is worth noting that the superiority of NOMA over orthogonal multiple access (OMA) was discussed by a considerable amount of literature \cite{Liu2017,Liu2023}. In light of this as well as the page limitations, further discussions on OMA-ISAC or OMA-FDSAC are skipped here and left as a potential direction for future work.

Given the NOMA-ISAC and NOMA-FDSAC frameworks, we intend to investigate the performance of communications as well as sensing. Particularly, we rely on the sensing rate for radar target sensing, which tells the information-theoretic limit on how much environmental information can be extracted from the reflected radar echoes, and employ the outage probability and ergodic communication rate to measure the performance of communications.

\section{Performance Analysis}
\subsection{Performance of Communications}
In this part, we investigate the communication performance of the NOMA-ISAC and NOMA-FDSAC systems. More specifically, we aim to derive exact expressions as well as high-SNR approximations of the outage probability and the ergodic communication rate, respectively.

Based on the NOMA technique, NU accomplishes the successive interference cancellation (SIC) procedure to detect FU's message and then remove the message from its observation, in a successive manner. Hence, the signal-to-interference-plus-noise ratio (SINR) of the SIC process for NU to decode FU's message is given by $\gamma_t^{\rm{sic}}=\frac{\mu_t p\left|h_{\rm{N}}\right|^2\alpha_{\rm{F}}}{\kappa_t\sigma_{\rm{c}}^2+\mu_t p\left|h_{\rm{N}}\right|^2\alpha_{\rm{N}}}$ for $t\in\left\{{\rm{i}},{\rm{f}}\right\}$, and the SNR for NU to decode its own message is given by
$\gamma_t^{\rm{N}}=\frac{\mu_t p\left|h_{\rm{N}}\right|^2\alpha_{\rm{N}}}{\kappa_t\sigma_{\rm{c}}^2}$, where the subscripts ``${\rm{i}}$'' and ``${\rm{f}}$'' indicate ISAC and FDSAC, respectively. It is worth noting that $\kappa_{\rm{i}}=\mu_{\rm{i}}=1$, $\kappa_{\rm{f}}=\kappa$, and $\mu_{\rm{f}}=\mu$. In contrast to NU, FU directly decodes its signal by treating NU's message as interference, and the corresponding SINR is given by
$\gamma_t^{\rm{F}}=\frac{\mu_t p\left|h_{\rm{F}}\right|^2\alpha_{\rm{F}}}{\kappa_t\sigma_{\rm{c}}^2+\mu_t p\left|h_{\rm{F}}\right|^2\alpha_{\rm{N}}}$ for $t\in\left\{{\rm{i}},{\rm{f}}\right\}$. Bearing in mind the above results, we analyze the communication performance of NOMA-FDSAC and NOMA-ISAC in two major steps which are illustrated in the sequel.
\subsubsection{Outage Probability}
At the first step, we investigate the outage performance of the two CUs. By definition, the outage probabilities of NU and FU for $t\in\left\{{\rm{i}},{\rm{f}}\right\}$ can be written as
\begin{align}
\mathcal{P}_t^{\rm{N}}&=1-\Pr\left(\gamma_t^{\rm{sic}}>\bar\gamma_t^{\rm{F}},\gamma_t^{\rm{N}}>\bar\gamma_t^{\rm{N}}\right),\label{Pout_NU}\\
\mathcal{P}_t^{\rm{F}}&=\Pr\left(\gamma_t^{\rm{F}}<\bar\gamma_t^{\rm{F}}\right),\label{Pout_FU}
\end{align}
respectively, where $\bar\gamma_t^{\rm{N}}=2^{\overline{\mathcal{R}}_{\rm{N}}/\kappa_t}-1$ and $\bar\gamma_t^{\rm{F}}=2^{\overline{\mathcal{R}}_{\rm{F}}/\kappa_t}-1$ with $\overline{\mathcal{R}}_{\rm{N}}$ and $\overline{\mathcal{R}}_{\rm{F}}$ being the target rates of NU and FU, respectively. The following theorem provides exact expressions for these OPs as well as their high-SNR approximations.
\vspace{-5pt}
\begin{theorem}\label{theorem_outage_probability}
When ${\alpha_{\rm{F}}\leq\bar\gamma_t^{\rm{F}}\alpha_{\rm{N}}}$, the OPs of NU and FU for $t\in\left\{{\rm{i}},{\rm{f}}\right\}$ are both given by $1$. When ${\alpha_{\rm{F}}>\bar\gamma_t^{\rm{F}}\alpha_{\rm{N}}}$, the OPs of NU and FU for $t\in\left\{{\rm{i}},{\rm{f}}\right\}$ are given by
\begin{subequations}\label{OP_Analytical_Expression}
\begin{align}
&\mathcal{P}_t^{\rm{N}}=\left(1-{\rm{e}}^{-\chi_{t,1}{\theta_t}{p^{-1}}}\right)\left(1-{\rm{e}}^{-\chi_{t,2}{\theta_t}{p^{-1}}}\right) ,\label{OP_Analytical_Expression1}\\
&\mathcal{P}_t^{\rm{F}}=
1-{\rm{e}}^{-\chi_{t,3}{\vartheta_t}{p^{-1}}},\label{OP_Analytical_Expression2}
\end{align}
\end{subequations}
respectively, where $\theta_t=\max\left\{\frac{\bar\gamma_t^{\rm{N}}}{ \alpha_{\rm{N}}},\vartheta_t\right\}$ with $\vartheta_t=\frac{\bar\gamma_t^{\rm{F}}}{\alpha_{\rm{F}}-\alpha_{\rm{N}}\bar\gamma_t^{\rm{F}}}$, and $\chi_{t,b}=\frac{\kappa_t\sigma_{\rm{c}}^2}{\mu_t\varrho_{b}}$ for $b\in\{1,2,3\}$ with $\varrho_3=\frac{\varrho_1\varrho_2}{\varrho_1+\varrho_2}$. Let ${\alpha_{\rm{F}}>\bar\gamma_t^{\rm{F}}\alpha_{\rm{N}}}$ and $p\rightarrow\infty$, the OPs satisfy:
\begin{align}\label{OP_Asym}
\mathcal{P}_t^{{\rm{N}}}
\approx{\chi_{t,1}\chi_{t,2}}{\theta_t^2}{p^{-2}},\quad
\mathcal{P}_t^{{\rm{F}}}
\approx{\chi_{t,3}{\vartheta_t}{p^{-1}}}.
\end{align}
\end{theorem}
\vspace{-5pt}
\begin{IEEEproof}
The proof is given in Appendix \ref{proof_theorem_outage_probability}.
\end{IEEEproof}
\vspace{-5pt}
\begin{remark}\label{remark_OP}
The results in \eqref{OP_Asym} suggest that when $\alpha_{\rm{F}}>\bar\gamma_t^{\rm{F}}\alpha_{\rm{N}}$, the diversity orders of NU and FU are given by $2$ and $1$, respectively, in both NOMA-ISAC and NOMA-FDSAC.
\end{remark}
\vspace{-5pt}
\subsubsection{Ergodic Rate}
At the second step, we discuss the ergodic communication rates of the two CUs. Particularly, the ergodic rates of NU and FU for $t\in\left\{{\rm{i}},{\rm{f}}\right\}$ are defined as
\begin{align}
\mathcal{R}_t^{\rm{N}}&={\mathbbmss{E}}\left\{{\kappa_t}\log_2\left(1+\gamma_t^{\rm{N}}\right)\right\},\\
\mathcal{R}_t^{\rm{F}}&={\mathbbmss{E}}\left\{{\kappa_t}\log_2\left(1+\gamma_t^{\rm{F}}\right)\right\},
\end{align}
respectively. Theorem \ref{theorem_ergodic_rate} provides exact expressions for these ECRs as well as their high-SNR approximations.
\vspace{-5pt}
\begin{theorem}\label{theorem_ergodic_rate}
The ECRs $\mathcal{R}_t^{\rm{N}}$ and $\mathcal{R}_t^{\rm{F}}$ are given by
\begin{subequations}\label{ECR_Analytical_Expression}
\begin{align}
&\mathcal{R}_t^{{\rm{N}}}=\frac{\kappa_t}{\ln{2}}\left(
\psi_{t,3}-\psi_{t,2}-\psi_{t,1}\right),\\
&\mathcal{R}_t^{{\rm{F}}}=\frac{\kappa_t}{\ln{2}}\psi_{t,3}-\frac{\kappa_t}{\ln{2}}
{{\rm{Ei}}\left({{-\chi_{t,3}}/{p}}\right)}{\rm{e}}^{{\frac{\chi_{t,3}}{p}}},
\end{align}
\end{subequations}
respectively, where ${\rm{Ei}}\left(\cdot\right)$ is the exponential integral and $\psi_{t,b}={{\rm{Ei}}\left(-\frac{\chi_{t,b}}{\alpha_{\rm{N}}p}\right)}{{\rm{e}}^{\frac{\chi_{t,b}}{\alpha_{\rm{N}}p}}}$. When $p\rightarrow\infty$, the ECRs satisfy:
\begin{subequations}\label{ECR_Asym}
\begin{align}
&\mathcal{R}_t^{{\rm{N}}}\approx\kappa_t\log_2{p}-\frac{\kappa_t\gamma}{\ln{2}}-\kappa_t
\log_2{\frac{\kappa_t\sigma_{\rm{c}}^2}{\mu_t{\alpha_{\rm{N}}}({\varrho_1+\varrho_2})}},\label{ECR_Asym_1}\\
&\mathcal{R}_t^{{\rm{F}}}\approx-\kappa_t\log_2{{\alpha_{\rm{N}}}},\label{ECR_Asym_2}
\end{align}
\end{subequations}
where $\gamma$ is the Euler–Mascheroni constant.
\end{theorem}
\vspace{-5pt}
\begin{IEEEproof}
The proof is given in Appendix \ref{proof_theorem_ergodic_rate}.
\end{IEEEproof}
\vspace{-5pt}
\begin{remark}\label{remark_ECR}
The results in \eqref{ECR_Asym} indicate that the high-SNR slopes of NU and FU in NOMA-FDSAC are given by $\kappa$ and $0$, respectively, whereas the high-SNR slopes of NU and FU in NOMA-ISAC are given by $1$ and $0$, respectively. This also means that the high-SNR slopes of the sum ECR of the CU pair, i.e., $\mathcal{R}_{t}^{\rm{c}}\triangleq\mathcal{R}_{t}^{{\rm{N}}}+\mathcal{R}_{t}^{{\rm{F}}}$ ($t\in\left\{{\rm{i}},{\rm{f}}\right\}$), in NOMA-FDSAC and NOMA-ISAC are given by $\kappa$ and $1$, respectively.
\end{remark}
\vspace{-5pt}

\subsection{Performance of Sensing}
Having characterized the performance of communications, we move on to analyzing the performance of sensing.

Let us first consider the NOMA-ISAC case. As mentioned previously, sensing the RTs is equivalent to extracting environmental information contained in the target response vector, i.e., $\mathbf{g}$ in \eqref{target_response_vector}, from the reflected radar echo signal matrix, i.e., ${\mathbf{Y}}_{\rm{s}}$ in \eqref{reflected_echo_signal_matrix}. To define the information-theoretic limit on how much environmental information can be extracted, we resort to the sensing rate for sensing performance evaluation. More specifically, the sensing rate is defined as the sensing mutual information (MI) per unit time and the sensing MI is defined as the MI between ${\mathbf{Y}}_{\rm{s}}$ and $\mathbf{g}$ conditioned on the radar waveform $\mathbf{x}$ \cite{Yuan2021,Tang2019}. Assuming that each waveform symbol in the ISAC system lasts $1$ unit time, we write the SR as follows:
\begin{align}
\mathcal{R}_{\rm{i}}^{\rm{s}}={L}^{-1}I\left({\mathbf{Y}}_{\rm{s}};{\mathbf{g}}|{\mathbf{x}}\right),
\end{align}
where $I\left(X;Y|Z\right)$ denotes the MI between $X$ and $Y$ conditioned on $Z$ \cite{Yuan2021,Tang2019}. Theorem \ref{theorem_sensing_rate} provides an exact expression for the SR as well as its high-SNR approximation.
\vspace{-5pt}
\begin{theorem}\label{theorem_sensing_rate}
The SR in the NOMA-ISAC system is given by
\begin{equation}\label{SR_Analytical_Expression}
\mathcal{R}_{\rm{i}}^{\rm{s}}={L^{-1}}\log_2\det\left({\mathbf{I}}_{M}+{pL}{\sigma_{\rm{s}}^{-2}}{\mathbf{R}}\right),
\end{equation}
where ${\mathbf{R}}={\mathbbmss{E}}\left\{{\mathbf{g}}{\mathbf{g}}^{\mathsf{H}}\right\}\in{\mathbbmss{C}}^{M\times M}$ is the correlation matrix of the target response vector $\mathbf{g}$. When $p\rightarrow\infty$, the SR satisfies:
\begin{equation}\label{SR_Asym}
\mathcal{R}_{\rm{i}}^{\rm{s}}\approx\frac{\mathsf{r}}{L}\log_2{p}+\frac{1}{L}\sum\nolimits_{a=1}^{\mathsf{r}}
\log_2\left(\frac{L\lambda_a}{\sigma_{\rm{s}}^2}\right),
\end{equation}
where ${\mathsf{r}}={\mathsf{rank}\left({\mathbf{R}}\right)}$ is the rank of matrix ${\mathbf{R}}$ and $\left\{\lambda_a\right\}_{a=1}^{\mathsf{r}}$ are the associated positive eigenvalues.
\end{theorem}
\vspace{-5pt}
\begin{IEEEproof}
The proof is given in Appendix \ref{proof_theorem_sensing_rate}.
\end{IEEEproof}
Following similar steps as those outlined in Appendix \ref{proof_theorem_sensing_rate}, we obtain the SR of the FDSAC system as follows.
\vspace{-5pt}
\begin{corollary}\label{theorem_sensing_rate_fdsac}
The SR of NOMA-FDSAC can be written as
\begin{equation}\label{SR_Analytical_Expression_fdsac}
\mathcal{R}_{\rm{f}}^{\rm{s}}=\frac{1-\kappa}{L}\log_2\det\left({\mathbf{I}}_{M}+\frac{\left(1-\mu\right)pL}{\left(1-\kappa\right)\sigma_{\rm{s}}^2}{\mathbf{R}}\right).
\end{equation}
When $p\rightarrow\infty$, the SR satisfies:
\begin{equation}\label{SR_Asym_fdsac}
\mathcal{R}_{\rm{f}}^{\rm{s}}\!\approx\!\frac{\mathsf{r}(1\!-\!\kappa)}{L}\log_2{p}
+\frac{1\!-\!\kappa}{L}\sum\nolimits_{a=1}^{{\mathsf{r}}}\!
\!\log_2{\frac{\left(1-\mu\right)\lambda_aL}{\left(1-\kappa\right)\sigma_{\rm{s}}^2}}.
\end{equation}
\end{corollary}
\vspace{-5pt}
\vspace{-5pt}
\begin{remark}\label{remark_sr}
The results in \eqref{SR_Asym} and \eqref{SR_Asym_fdsac} indicate that the high-SNR slopes of $\mathcal{R}_{\rm{i}}^{\rm{s}}$ and $\mathcal{R}_{\rm{f}}^{\rm{s}}$ are given by $\frac{\mathsf{r}}{L}$ and $\frac{\left(1-\kappa\right)\mathsf{r}}{L}$, respectively.
\end{remark}
\vspace{-5pt}
Last but not least, although the sensing rate defines the information-theoretic limit on how much environmental information can be extracted, how to extract this information relies on specific estimation or detection algorithms \cite{Liu2022}. Yet, this is beyond the scope of this paper.

\subsection{Summary of the Analytical Results}
After completing all analyses, we summarize the results related to diversity order and high-SNR slope in Table \ref{table1} for ease of reference. Since $\kappa\in[0,1]$, it is observed that NOMA-ISAC achieves larger high-SNR slopes than NOMA-FDSAC in terms of the SR as well as the sum ECR of the CU pair. This reflects that NOMA-ISAC can provide more degrees of freedom for both communications and sensing than NOMA-FDSAC. In fact, this superiority mainly originates from ISAC's integrated exploitation of spectrum and power resources \cite{Liu2022}.
\begin{table}[htbp]
\centering
\begin{tabular}{|c|cc|cc|c|c|}
\hline
\multirow{2}{*}{System} & \multicolumn{2}{c|}{N}     & \multicolumn{2}{c|}{F}     & CU Pair & Sensing \\ \cline{2-7}
                        & \multicolumn{1}{c|}{$\mathcal{D}$} & $\mathcal{S}$ & \multicolumn{1}{c|}{$\mathcal{D}$} & $\mathcal{S}$ & $\mathcal{S}$       & $\mathcal{S}$       \\ \hline
NOMA-FDSAC                   & \multicolumn{1}{c|}{2} & $\kappa$ & \multicolumn{1}{c|}{1} & 0 & $\kappa$       & $\frac{\left(1-\kappa\right)\mathsf{r}}{L}$       \\ \hline
NOMA-ISAC                    & \multicolumn{1}{c|}{2} & 1 & \multicolumn{1}{c|}{1} & 0 & 1       & $\frac{\mathsf{r}}{L}$       \\ \hline
\end{tabular}
\caption{Diversity Order ($\mathcal{D}$) and High-SNR Slope ($\mathcal{S}$)}
\label{table1}
\end{table}

\section{Rate Region Characterization}
The previous part of this paper has analyzed the basic performances of the NOMA-ISAC and NOMA-FDSAC systems. It is now necessary to characterize the sensing-communication rate region of these two systems as well as providing a full and complete comparison between them.
\subsection{Achievable Sensing-Communication Rate Region}
In this subsection, we aim to characterize the achievable sensing-communication rate region of the NOMA-ISAC and NOMA-FDSAC systems. To this end, let ${\mathcal{R}}^{\rm{s}}$ and ${\mathcal{R}}^{\rm{c}}$ denote the achievable SR and sum ECR, respectively. Then, the rate regions achieved by ISAC and FDSAC are given by
\begin{align}
&\mathcal{C}_{\rm{isac}}=\left\{\left({\mathcal{R}}^{\rm{s}},{\mathcal{R}}^{\rm{c}}\right)|{\mathcal{R}}^{\rm{s}}\in\left[0,\mathcal{R}_{\rm{i}}^{\rm{s}}\right],
{\mathcal{R}}^{\rm{c}}\in\left[0,\mathcal{R}_{\rm{i}}^{\rm{c}}\right]\right\},\label{Rate_Regio_ISAC}\\
&\mathcal{C}_{\rm{fdsac}}=\left\{\left({\mathcal{R}}^{\rm{s}},{\mathcal{R}}^{\rm{c}}\right)\left|
\begin{aligned}
&{\mathcal{R}}^{\rm{s}}\in\left[0,\mathcal{R}_{\rm{f}}^{\rm{s}}\right],
{\mathcal{R}}^{\rm{c}}\in\left[0,\mathcal{R}_{\rm{f}}^{\rm{c}}\right],\\
&\kappa\in\left[0,1\right],\mu\in\left[0,1\right]
\end{aligned}
\right.\right\},\label{Rate_Regio_FDSAC}
\end{align}
respectively. This subsection has presented the rate regions of NOMA-ISAC and NOMA-FDSAC. The next part of this section will estimate the relationship between $\mathcal{C}_{\rm{isac}}$ and $\mathcal{C}_{\rm{fdsac}}$.
\subsection{Comparison Between NOMA-ISAC and NOMA-FDSAC}
Let us introduce some preliminary results that will be useful in discussing the relationship between $\mathcal{C}_{\rm{isac}}$ and $\mathcal{C}_{\rm{fdsac}}$.
\vspace{-5pt}
\begin{lemma}\label{lemma_ECR_Comparision}
Given $\kappa\in\left[0,1\right]$ and $\mu\in\left[0,1\right]$. The sum ECRs of the CU pair in NOMA-ISAC and NOMA-FDSAC satisfy $\mathcal{R}_{\rm{i}}^{\rm{c}}\geq\mathcal{R}_{\rm{f}}^{\rm{c}}$, where the equality holds for $\kappa=\mu=1$.
\end{lemma}
\vspace{-5pt}
\begin{IEEEproof}
The proof is given in Appendix \ref{proof_lemma_ECR_Comparision}.
\end{IEEEproof}
\vspace{-5pt}
\begin{lemma}\label{lemma_SR_Comparision}
Given $\kappa\in\left[0,1\right]$ and $\mu\in\left[0,1\right]$. The SRs of the NOMA-ISAC and NOMA-FDSAC systems satisfy $\mathcal{R}_{\rm{i}}^{\rm{s}}\geq\mathcal{R}_{\rm{f}}^{\rm{s}}$, where the equality holds for $\kappa=\mu=0$.
\end{lemma}
\vspace{-5pt}
\begin{IEEEproof}
The proof is given in Appendix \ref{proof_lemma_ECR_Comparision}.
\end{IEEEproof}
Leveraging the above two lemmas, we present the relationship between $\mathcal{C}_{\rm{isac}}$ and $\mathcal{C}_{\rm{fdsac}}$ as follows.
\vspace{-5pt}
\begin{theorem}\label{theorem_Rate_Region_Comparision}
The achievable rate regions of NOMA-ISAC and NOMA-FDSAC satisfy $\mathcal{C}_{\rm{fdsac}}\subseteq\mathcal{C}_{\rm{isac}}$.
\end{theorem}
\vspace{-5pt}
\begin{IEEEproof}
By Lemmas \ref{lemma_ECR_Comparision} and \ref{lemma_SR_Comparision}, for any rate pair $\left({\mathcal{R}}^{\rm{s}},{\mathcal{R}}^{\rm{c}}\right)\in\mathcal{C}_{\rm{fdsac}}$, it satisfies ${\mathcal{R}}^{\rm{s}}\leq\mathcal{R}_{\rm{f}}^{\rm{s}}\leq\mathcal{R}_{\rm{i}}^{\rm{s}}$ and ${\mathcal{R}}^{\rm{c}}\leq\mathcal{R}_{\rm{f}}^{\rm{c}}\leq\mathcal{R}_{\rm{i}}^{\rm{c}}$, which yields $\left({\mathcal{R}}^{\rm{s}},{\mathcal{R}}^{\rm{c}}\right)\in\mathcal{C}_{\rm{isac}}$. It follows that $\mathcal{C}_{\rm{fdsac}}\subseteq\mathcal{C}_{\rm{isac}}$.
\end{IEEEproof}
\vspace{-5pt}
\begin{remark}
The results in Theorem \ref{theorem_Rate_Region_Comparision} suggest that the rate region achieved by FDSAC can be entirely covered by that achieved by ISAC, which highlights the superiority of ISAC.
\end{remark}
\vspace{-5pt}

\section{Numerical Results}
In this section, computer simulations will be employed to demonstrate the performance of the considered NOMA-ISAC system and also verify the accuracy of the developed analytical results. The parameters used for simulation are listed as follows: $M=\mathsf{r}=8$, $\left\{\lambda_a\right\}_{a=1}^{\mathsf{r}}=\left\{5,3,3.5,2.5,1.5,2,1,0.5\right\}$, $L=30$, $\varrho_{1}=0.9$, $\varrho_{2}=0.2$, $\alpha_{\rm{F}}=0.8$, and $\alpha_{\rm{N}}=0.2$.

\begin{figure}[!t]
    \centering
    \subfigbottomskip=0pt
	\subfigcapskip=-5pt
\setlength{\abovecaptionskip}{0pt}
    \subfigure[Outage probability.]
    {
        \includegraphics[height=0.17\textwidth]{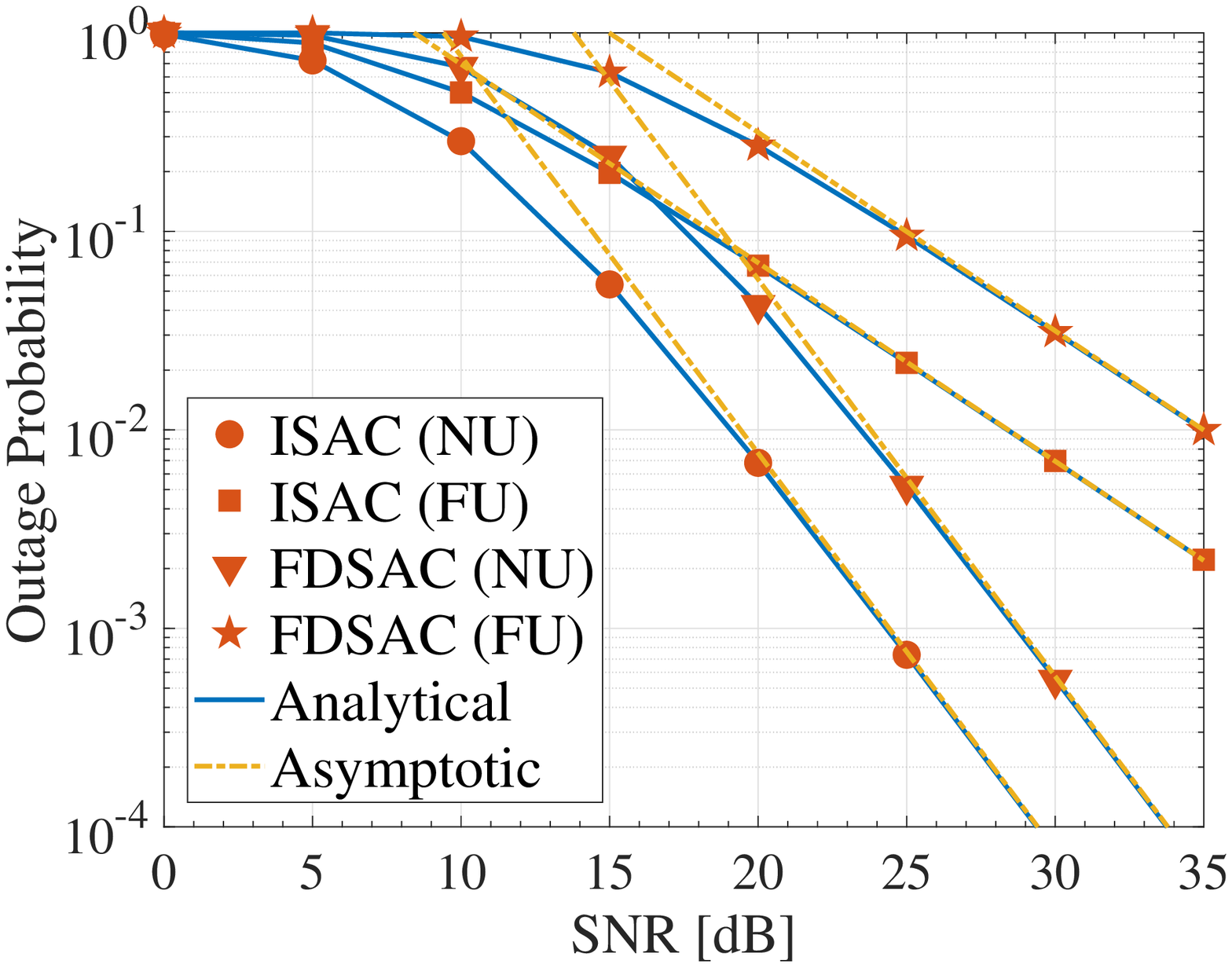}
	   \label{fig1a}	
    }
   \subfigure[Ergodic communication rate.]
    {
        \includegraphics[height=0.17\textwidth]{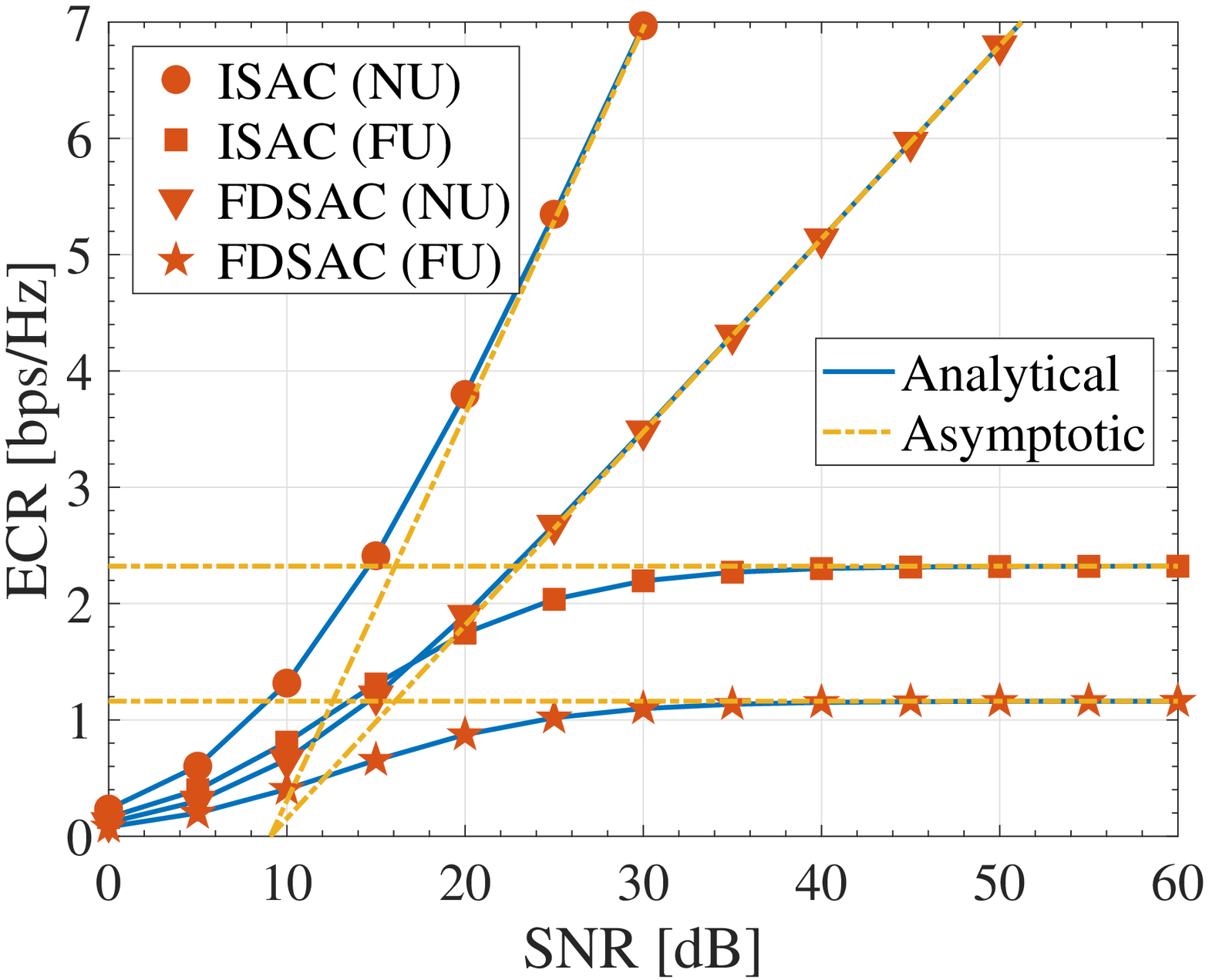}
	   \label{fig1b}	
    }
\caption{Performance of communications. $\overline{\mathcal{R}}_{\rm{N}}=\overline{\mathcal{R}}_{\rm{F}}=0.8$ bps/Hz and $\kappa=\mu=0.5$.}
    \label{Figure1}
    \vspace{-5pt}
\end{figure}

\begin{figure}[!t]
\centering
\setlength{\abovecaptionskip}{0pt}
\includegraphics[height=0.2\textwidth]{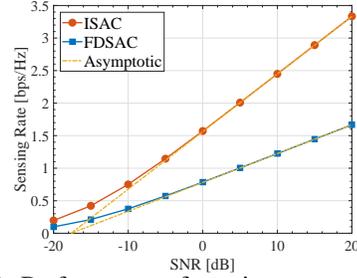}
\caption{Performance of sensing. $\kappa=\mu=0.5$.}
\label{Figure2}
\vspace{-5pt}
\end{figure}

{\figurename} {\ref{fig1a}} and {\figurename} {\ref{fig1b}} plot the OP and ECR versus the SNR $p$, respectively, where the simulation results (denoted by symbols) match the analytical results developed at \eqref{OP_Analytical_Expression} and \eqref{ECR_Analytical_Expression}. It is worth pointing out that the asymptotic results derived at \eqref{OP_Asym} and \eqref{ECR_Asym} can track the provided simulation results in {\figurename} {\ref{fig1a}} and {\figurename} {\ref{fig1b}} accurately in the high-SNR regime. As can be observed from {\figurename} {\ref{fig1a}}, although FDSAC can provide the same diversity order as ISAC, ISAC yields lower OPs for both NU and FU than FDSAC. Moreover, it can be seen from {\figurename} {\ref{fig1b}} that ISAC outperforms FDSAC in terms of NU's ECR, NU's high-SNR slope, and FU's ECR, which is consistent with the conclusions drawn in Remark \ref{remark_ECR} and Appendix \ref{proof_lemma_ECR_Comparision}. In {\figurename} {\ref{Figure2}}, the sensing rate is shown as a function of the SNR $p$, where the asymptotic results are calculated by \eqref{SR_Asym} and \eqref{SR_Asym_fdsac}. As can be seen from this graph, the asymptotic results track the provided simulation results accurately in the high-SNR regime. From the data in {\figurename} {\ref{Figure2}}, it is apparent that the ISAC outperforms FDSAC in terms of both sensing rate and high-SNR slope, which agrees with the results presented in Remark \ref{remark_sr} and Theorem \ref{theorem_Rate_Region_Comparision}. Taken together, the above numerical studies reveal that NOMA-ISAC is capable of providing better performance of communications as well as sensing than NOMA-FDSAC.

\begin{figure}[!t]
\centering
\setlength{\abovecaptionskip}{0pt}
\includegraphics[height=0.2\textwidth]{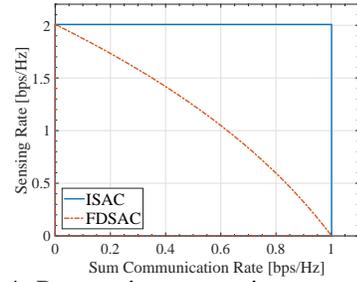}
\caption{Rate region comparison. $p=5$ dB.}
\label{Figure3}
\vspace{-5pt}
\end{figure}

{\figurename} {\ref{Figure3}} compares the sensing-communication rate regions achieved by the ISAC system (presented in \eqref{Rate_Regio_ISAC}) and the baseline FDSAC system (presented in \eqref{Rate_Regio_FDSAC}). In particular, the rate region of the FDSAC system is plotted by changing the bandwidth allocation factor $\kappa$ as well as the power allocation factor $\mu$ in \eqref{Rate_Regio_FDSAC} from $0$ to $1$. What stands out in this graph is that the rate region achieved by FDSAC is completely included in that achieved by ISAC, which, therefore, verifies the correctness of Theorem \ref{theorem_Rate_Region_Comparision}. This, together with the numerical results in {\figurename} {\ref{Figure1}} and {\figurename} {\ref{Figure2}}, highlights the superiority of NOMA-ISAC over the conventional NOMA-FDSAC scheme.

\section{Conclusion}
In this letter, we have analyzed the performance and characterized the achievable sensing-communication rate region of NOMA-ISAC systems. Simulation results have verified the accuracy of the developed analytical results. Theoretical analyses have demonstrated that NOMA-ISAC is capable of providing more degrees of freedom as well as achieving a larger rate region compared with NOMA-FDSAC.

\begin{appendices}
\section{Proof of Theorem \ref{theorem_outage_probability}}\label{proof_theorem_outage_probability}
The probability density functions (PDFs) of $\left|h_{\rm{N}}\right|^2$ and $\left|h_{\rm{F}}\right|^2$ can be written as $f_{\rm{N}}(x)=\frac{1}{\varrho_1}{\rm{e}}^{-\frac{x}{\varrho_1}}+\frac{1}{\varrho_2}{\rm{e}}^{-\frac{x}{\varrho_2}}-\frac{1}{\varrho}{\rm{e}}^{-\frac{x}{\varrho}}$ and $f_{\rm{F}}(x)=\frac{1}{\varrho}{\rm{e}}^{-\frac{x}{\varrho}}$, respectively. The cumulative distribution functions (CDFs) of $\left|h_{\rm{N}}\right|^2$ and $\left|h_{\rm{F}}\right|^2$ can be written as $F_{\rm{N}}\left(x\right)=(1-{\rm{e}}^{-{x}/{\varrho_1}})(1-{\rm{e}}^{-{x}/{\varrho_2}})$ and $F_{\rm{F}}\left(x\right)=1-{\rm{e}}^{-{x}/{\varrho}}$, respectively. Note that $\gamma_t^{\rm{sic}}=\frac{\mu_t p\left|h_{\rm{N}}\right|^2\alpha_{\rm{F}}}{\kappa_t\sigma_{\rm{c}}^2+\mu_t p\left|h_{\rm{N}}\right|^2\alpha_{\rm{N}}}$. Hence, when ${\alpha_{\rm{F}}<\bar\gamma_t^{\rm{F}}\alpha_{\rm{N}}}$, we get $\Pr\left(\gamma_t^{\rm{sic}}>\bar\gamma_t^{\rm{F}}\right)=\Pr\left(\left|h_{\rm{N}}\right|<0\right)=1$, whereas when ${\alpha_{\rm{F}}=\bar\gamma_t^{\rm{F}}\alpha_{\rm{N}}}$, we get $\Pr\left(\gamma_t^{\rm{sic}}>\bar\gamma_t^{\rm{F}}\right)=\Pr\left(0>\bar\gamma_t^{\rm{F}}\right)=1$. As a result, the OP of NU is given by ${\mathcal{P}}_t^{\rm{N}}=1$ when ${\alpha_{\rm{F}}<\bar\gamma_t^{\rm{F}}\alpha_{\rm{N}}}$. Moreover, by the definition of ${\mathcal{P}}_t^{\rm{N}}$, the OP of NU can be calculated as ${\mathcal{P}}_t^{\rm{N}}=\Pr\left(\left|h_{\rm{N}}\right|^2>\frac{\kappa_t\bar\gamma_t^{\rm{F}}\sigma_{\rm{c}}^2}
{\mu_tp\varrho_{\rm{N}}^2\alpha_{\rm{F}}-\mu_tp\varrho_{\rm{N}}^2\alpha_{\rm{N}}\bar\gamma_t^{\rm{F}}},\left|h_{\rm{N}}\right|^2>\frac{\kappa_t\bar\gamma_t^{\rm{N}}\sigma_{\rm{c}}^2}{\mu_t\alpha_{\rm{N}}p\varrho_{\rm{N}}^2}\right)$. Leveraging the PDF of $\left|h_{\rm{N}}\right|^2$, we get \eqref{OP_Analytical_Expression1} after some basic mathematical manipulations. Then, following similar steps as listed above, we also get \eqref{OP_Analytical_Expression2}. As $p\rightarrow\infty$, we obtain the asymptotic results in \eqref{OP_Asym} based on the fact of $\lim_{x\rightarrow0}\left(1-{\rm{e}}^{-x}\right)\approx x$.
\section{Proof of Theorem \ref{theorem_ergodic_rate}}\label{proof_theorem_ergodic_rate}
Inserting the obtained PDF expressions into the expressions of the ECRs and calculating the resultant integrals with the aid of \cite[Eq. (4.337.2)]{Ryzhik2007}, we obtain the results in \eqref{ECR_Analytical_Expression}. By continuously using the fact of $\lim_{x\rightarrow 0}{\rm{Ei}}(-x)\approx\gamma+\ln{x}$ \cite[Eq. (8.214.2)]{Ryzhik2007}, we get the approximated results in \eqref{ECR_Asym}.
\section{Proof of Theorem \ref{theorem_sensing_rate}}\label{proof_theorem_sensing_rate}
Define $\mathbf{y}={\mathsf{vec}}\left({\mathbf{Y}}_{\rm{s}}^{\mathsf{T}}\right)\in{\mathbbmss{C}}^{LM\times 1}$, $\mathbf{n}={\mathsf{vec}}\left({\mathbf{N}}_{\rm{s}}^{\mathsf{T}}\right)\sim{\mathcal{CN}}\left(\mathbf{0},\sigma_{\rm{s}}^2\mathbf{I}_{LM}\right)$, and $\mathbf{X}=\mathbf{I}_{M}\otimes\mathbf{x}\in{\mathbbmss{C}}^{LM\times M}$, where the symbol $\otimes$ represents the Kronecker product. Then, we can rewrite \eqref{reflected_echo_signal_matrix} as $\mathbf{y}=\mathbf{X}\mathbf{g}+\mathbf{n}$. Therefore, the sensing MI satisfies
$I\left({\mathbf{Y}}_{\rm{s}};{\mathbf{g}}|{\mathbf{x}}\right)=I\left({\mathbf{y}};{\mathbf{g}}|{\mathbf{X}}\right)=\log_2\det({\mathbf{I}}_{LM}
+{\sigma_{\rm{s}}^{-2}}{\mathbf{X}}\mathbf{R}{\mathbf{X}}^{\mathsf{H}})$, which together with the Sylvester's identity, yields $I\left({\mathbf{Y}}_{\rm{s}};{\mathbf{g}}|{\mathbf{x}}\right)=\log_2\det({\mathbf{I}}_{M}
+{\sigma_{\rm{s}}^{-2}}\mathbf{R}{\mathbf{X}}^{\mathsf{H}}{\mathbf{X}})$. We note that ${\mathbf{X}}^{\mathsf{H}}{\mathbf{X}}={\mathbf{x}}^{\mathsf{H}}{\mathbf{x}}{\mathbf{I}}_M$. Based on \eqref{dual_function_signal_matrix} and \eqref{normalized_signal_matrix}, we get ${\mathbf{x}}^{\mathsf{H}}{\mathbf{x}}=pL$ and the results in \eqref{SR_Analytical_Expression} follow immediately. Using the eigendecomposition of $\mathbf{R}$, the sensing rate in \eqref{SR_Analytical_Expression} can be simplified to $\mathcal{R}_{\rm{i}}^{\rm{s}}=\sum\nolimits_{a=1}^{\mathsf{r}}\log_2(1+{pL}{\sigma_{\rm{s}}^{-2}}\lambda_a)$. By setting $p\rightarrow\infty$, the results in \eqref{SR_Asym} follow immediately.
\section{Proof of Lemmas \ref{lemma_ECR_Comparision} and \ref{lemma_SR_Comparision}}\label{proof_lemma_ECR_Comparision}
Before starting the proof, we introduce the following preliminary results that will be used.
\vspace{-5pt}
\begin{lemma}\label{Lemma_Help1}
Given $c\geq0$. The function $g_c(x)=\ln\left(x+c\right)+\frac{c}{x+c}$ increases with $x\in\left[0,\infty\right)$ monotonically.
\end{lemma}
\vspace{-5pt}
\begin{IEEEproof}
The derivative of $g_c(x)$ with respect to $x$ is given by $\frac{{\rm{d}}}{{\rm{d}}x}g_c(x)=\frac{x}{(x+c)^2}\geq0$, where the equality only holds for $x=0$ and $c>0$. Thus, this lemma is proved.
\end{IEEEproof}
\vspace{-5pt}
\begin{lemma}\label{Lemma_Help2}
Given $a>0$ and $b\geq0$. The function $f(x)=x\ln(1+\frac{a}{x+b})$ increases with $x\in\left[0,1\right]$ monotonically.
\end{lemma}
\vspace{-5pt}
\begin{IEEEproof}
The derivative of $f(x)$ with respect to $x$ is given by $\frac{{\rm{d}}}{{\rm{d}}x}f(x)=g_x(a+b)-g_x(b)$, which together with the fact of $a+b>b\geq0$ and Lemma \ref{Lemma_Help1}, suggests that $\frac{{\rm{d}}}{{\rm{d}}x}f(x)>0$ holds for $x\geq0$. Thus, this lemma is proved.
\end{IEEEproof}
We now intend to prove Lemmas \ref{lemma_ECR_Comparision} and \ref{lemma_SR_Comparision} with the aid of Lemma \ref{Lemma_Help2}. Particularly, using Lemma \ref{Lemma_Help2} and the fact of $\kappa\in[0,1]$, we can get $\kappa\log_2(1+\frac{ y_1}{\kappa+y_2})\leq\log_2(1+\frac{y_1}{1+y_2})$ ($y_1>0$ and $y_2\geq0$), where the equality holds for $\kappa=1$. Moreover, using the monotonicity of $q(t)=\frac{y_1t}{1+y_2t}$ for $t\geq0$ and the fact of $\mu\in[0,1]$, we obtain $\log_2(1+\frac{y_1}{1+y_2})\leq\log_2(1+\frac{y_1/\mu}{1+y_2/\mu})$, where the equality holds for $\mu=1$. Taken together, we get $\kappa\log_2(1+\frac{ y_1}{\kappa+y_2})\leq\log_2(1+\frac{y_1/\mu}{1+y_2/\mu})$.
By setting $y_1={\mu p\left|h_{\rm{F}}\right|^2\alpha_{\rm{F}}}/{\sigma_{\rm{c}}^2}$ and $y_2={\mu p\left|h_{\rm{F}}\right|^2\alpha_{\rm{N}}}/{\sigma_{\rm{c}}^2}$ and taking the expectation with respect to $\left|h_{\rm{F}}\right|^2$ on each side of this inequality, we have $\mathcal{R}_{\rm{f}}^{{\rm{F}}}\leq\mathcal{R}_{\rm{i}}^{{\rm{F}}}$. Besides, by setting $y_1={\mu p\left|h_{\rm{N}}\right|^2\alpha_{\rm{N}}}/{\sigma_{\rm{c}}^2}$ and $y_2=0$, we have $\mathcal{R}_{\rm{f}}^{{\rm{N}}}\leq\mathcal{R}_{\rm{i}}^{{\rm{N}}}$. In summary, we obtain $\mathcal{R}_{\rm{i}}^{\rm{c}}=\mathcal{R}_{\rm{i}}^{\rm{N}}+\mathcal{R}_{\rm{i}}^{\rm{F}}\geq\mathcal{R}_{\rm{f}}^{\rm{N}}+\mathcal{R}_{\rm{f}}^{\rm{F}}=\mathcal{R}_{\rm{f}}^{\rm{c}}$, and thus Lemma \ref{lemma_ECR_Comparision} is proved. Following the way we obtain $\mathcal{R}_{\rm{f}}^{{\rm{N}}}\leq\mathcal{R}_{\rm{i}}^{{\rm{N}}}$, we can get $\mathcal{R}_{\rm{i}}^{\rm{s}}\geq\mathcal{R}_{\rm{f}}^{\rm{s}}$. Thus, Lemma \ref{lemma_SR_Comparision} is proved.
\end{appendices}

\end{document}